# Code Generator Composition for Model-Driven Engineering of Robotics Component & Connector Systems

Jan Oliver Ringert[1,2*], Alexander Roth[1], Bernhard Rumpe[1],
Andreas Wortmann[1]

[1] Software Engineering
RWTH Aachen University
http://www.se-rwth.de/
[2] School of Computer Science
Tel Aviv University
http://www.cs.tau.ac.il/

**Abstract.** Engineering software for robotics applications requires multi-domain and application-specific solutions. Model-driven engineering and modeling language integration provide means for developing specialized, yet reusable models of robotics software architectures. Code generators transform these platform independent models into executable code specific to robotic platforms. Generative software engineering for multi-domain applications requires not only the integration of modeling languages but also the integration of validation mechanisms and code generators. In this paper we sketch a conceptual model for code generator composition and show an instantiation of this model in the MontiArc-Automaton framework. MontiArcAutomaton allows modeling software architectures as component and connector models with different component behavior modeling languages. Effective means for code generator integration are a necessity for the post hoc integration of application-specific languages in model-based robotics software engineering.

## 1 Introduction

Software engineering for robotic systems is inherently complex due to the heterogeneity of the systems and their challenges from various domains (e.g., navigation, sensor fusion, manipulation). Thus, robotics software is usually developed by teams of domain experts with different views and understanding of the systems functionality. This leads to hardly reusable software limited to specific platforms [8, 17]. To enable the reuse of functionality and subsystems, the structuring and composition mechanisms of component-based software engineering have been applied to robotics software [4, 12]. These approaches are mainly based on the exchange of source code components and thus tied to specific platforms and general-purpose programming languages (GPL). The application of

* J. O. Ringert acknowledges support from a postdoctoral Minerva Fellowship, funded by the German Federal Ministry for Education and Research.

GPLs often does not reflect the problems from heterogeneous domains faced in the development of robotics systems.

Model-driven engineering (MDE) is an approach to reduce the *conceptual gap* [5] between problem domains and software engineering. Models allow domain-specific software descriptions reflecting the heterogeneity of the developed system and its concerns. In combination with powerful code generators models may serve as primary development artifacts which increases the software's comprehensibility and reuse on different platforms.

We have combined MDE and software language engineering based approaches with concepts from generative software development in a versatile framework for robotics applications development. MontiArcAutomaton [14,15] is an extensible framework that allows to model robotics applications as hierarchically composable components with well-defined interfaces that embed problem specific modeling languages for component behavior. MontiArcAutomaton comprises powerful code generation facilities for the transformation of models into executable code for various robotics target platforms.

Language integration in MontiArcAutomaton is enabled by the MontiCore domain-specific language workbench [10]. MontiCore provides comprehensive language composition mechanisms supported by its symbol table and code generation frameworks [16,21]. We have presented the language composition mechanisms used by MontiArcAutomaton in [11].

The easy integration of modeling languages demands for integration mechanisms of corresponding code generators. Challenges are the coordination of multiple code generators each responsible for specific models or parts of models. This includes the selection of code generators supporting a common target platform, to handle language restrictions a code generator might impose, and to propagate necessary *generation context information* between generators. Integrating code generators should require no modification of the participating generators.

In this paper we sketch a conceptual model for code generator composition and show its instantiation in the MontiArcAutomaton framework. We introduce MontiArcAutomaton in Sect. 2 and state and illustrate the problem of code generator composition in Sect. 3. Section 4 describes our solution and Sect. 5 describes an implementation in MontiArcAutomaton. We discuss related work in Sect. 6 and future work in Sect. 7. Section 8 concludes this contribution.

## 2   MontiArcAutomaton

MontiArcAutomaton [14,15] is an extensible modeling language and framework for the generative model-driven engineering of robotics applications. The modeling language MontiArcAutomaton is a component and connector (C&C) architecture description language (ADL) [19] which extends the ADL MontiArc [7] with component behavior modeling. The logical architecture of robotics applications is described as the hierarchical composition of components that encapsulate the system's functionality. Components are either atomic or composed: atomic components define behavior via an embedded behavior modeling language or a component implementation in a general-purpose programming language. The behavior of composed components emerges from the subcomponents and their interaction. Components interact by sending messages via directed connectors that connect typed input and output ports of components. Types of ports are either defined via class diagrams or Java classes. Communication in MontiArc-Automaton is based on the Focus [3] framework for interactive distributed systems and supports different timing paradigms.

The concept of encapsulation from C&C ADLs allows not only a logically distributed development and a physically distributed computation model but also the composition of component behaviors independent of their behavior description. MontiArcAutomaton exploits the C&C encapsulation mechanism and allows the embedding of arbitrary modeling languages into components for providing the most suitable behavior description language per component.

We have developed MontiArcAutomaton using the domain-specific language workbench MontiCore [10] and its language integration mechanisms. The concrete and abstract syntax of a textual MontiCore modeling language is defined in an extended context free grammar format. From these grammars, MontiCore generates infrastructure to parse models of this language into their abstract syntax trees (ASTs). Checks of the well-formedness of models of a language, called *context conditions*, are implemented in Java [21]. MontiCore languages are textual modeling languages. An integration with the Eclipse Modeling Framework allows also the development of graphical editors for editing MontiArcAutomaton models.[3]

MontiCore supports language embedding, language extension, and language aggregation [11,21] to compose new languages from existing ones. These modular language composition mechanisms are supported by a sophisticated symbol table framework that enables the definition and adaptation of language symbols for integrating information and checking context conditions rules across embedded and imported models. MontiCore allows the easy development of code generators using the FreeMarker[4] template engine to process abstract syntax trees and code templates written in a target language [13,16].

In previous works we have developed the MontiArcAutomaton modeling language with embedded I/O$^\omega$ automata and I/O tables [15]. Various code generators allow the deployment of MontiArcAutomaton models to different robotics platforms [14,15]. With the integration of additional languages to model component behavior the post hoc composition of code generators has become a prevalent challenge.

## 3    Problem Statement and Example

MontiArcAutomaton allows to embed *application-specific* behavior modeling languages into components to facilitate the development of flexible, reusable, yet specific robotics applications. While the ability to use specific behavior modeling languages allows to develop specific applications, the encapsulation of models in components with well-defined and stable interfaces allows to modify component internals easily, e.g., to replace the specific behavior modeling language, while retaining a stable architecture.

---

[3] Video of an editor for synchronous graphical and textual editing of MontiArc-Automaton models: http://www.monticore.de/robotics/
[4] Website of the FreeMarker Java template engine: http://freemarker.org/

Engineering C&C applications with the flexibility of arbitrary embedded behavior modeling languages demands for approaches to generate code from heterogeneous models. As languages and code generators can be integrated into MontiArcAutomaton post hoc, code generators have to be composable to allow black-box integration. Each composable code generator produces only parts of the overall generated software system. A framework to support code generator composition has to provide a mechanism to configure C&C applications with different code generators. Realizing composition of code generators requires support for code generator reuse, the ability to handle code generators that are agnostic of any component structure specifics (e.g., how port or connectors work), and dependency management between different code generators.

## 3.1 Example

A software engineer is responsible for the development of a controller for a robotic arm. The robot assists a physically disabled person in a kitchen environment to operate a toaster. The robot is supposed to place bread in a toaster, operate the toaster, and deliver the toast to a nearby plate. The software engineer models the architecture and controller behavior platform independently using MontiArcAutomaton with embedded I/O$^\omega$ automata and RobotArm (RA) programs. The latter describe motion of the arm in terms of defined locations and gripper commands.[5] The engineer embeds the existing language RA into the MontiArcAutomaton framework using the language integration mechanisms of MontiCore. The software architecture of the robot is depicted in Fig. 1. The component `Controller` receives distances and toast color from attached sensors. The I/O$^\omega$ automaton modeling the behavior of `Controller` translates these inputs into commands for the `ToasterController`, which starts and stops the actual toaster, and the component `ArmController`, which actuates the robotic arm to pick up and deliver toast. The behavior of component `ArmController` is modeled as a set of RA programs.

To generate executable code from the architecture, the software engineer has to provide a code generator for the embedded RA language which translates RA commands into code for the target platform. This code generator can be selected from a library of existing code generators or newly developed. Finally, the generator has to be integrated into the framework, such that it is executed whenever a component with RA programs is processed.

---

[5] A video of the robotic arm: `http://www.monticore.de/robotics`

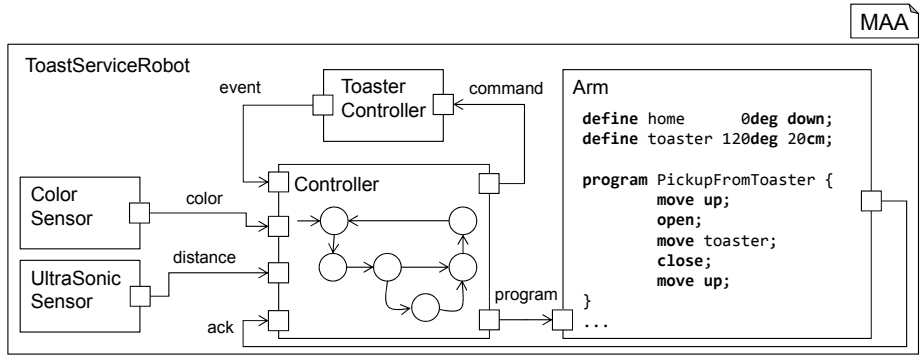

Fig. 1: Architecture of the `ToastServiceRobot` with embedded RobotArm programs.

## 4  Code Generator Composition

In this section we propose an approach to code generator composition on a conceptual level. First, we describe code generator interfaces that support generator composition as motivated in Sect. 3. Second, we sketch the process of code generator composition and execution of the composed generators using information from code generator interfaces.

To achieve generator composition, each code generator explicates all information necessary within an interface. This interface is used during code generator orchestration to configure and execute the code generator. Definition 1 lists the elements of a code generator interface.

**Definition 1 (Code Generator Interface).** A code generator interface contains the following elements:

1. Input language: The language or language fragments the generator processes.
2. Input language constraints: A generator may restrict the processable models via generator-specific context conditions.
3. Output representation: The output representation states the language and format of the output.
4. Execution information: Defines how a generator is executed.
5. Artifact dependencies: A generator may produce code that depends on external libraries, runtimes, or code produced by another code generator. Such artifact dependencies have to be explicitly stated in order to satisfy dependencies of generated artifacts.
6. Generation context information: Additional information provided or required at generation time.

Multiple generators are composed to generate code for models of the software of a robotic system. The composition of code generators is described in an *application configuration* model which contains a selection of all code generators involved. It may also contain a configuration of generation context information for code generators. Composing generators according to a configuration model requires the orchestration of all selected code generators. Such an orchestration requires (a) to check that all required information is provided and (b) to compute an execution order of the code generators.

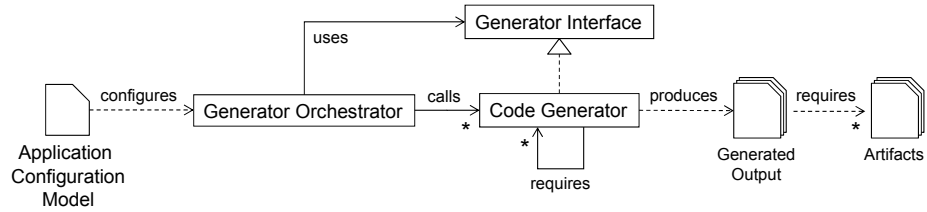

Fig. 2: Overview of code generator composition with generator interfaces.

If for each code generator all required generation context information is provided by the selected code generators and an execution order can be computed, then the code generator composition can be performed. However, the execution order of the code generators is influenced by the dependencies described by the generation context information. There are two types of dependencies. First, a code generator may require generation context information from another code generator. Second, a code generator may use the output of another code generator. Both types of dependencies imply that the code generator providing required information or output is executed first. However, in some cases it is possible that an execution order cannot be computed. In this case the code generators cannot be composed.

Our concept of code generator composition is presented in Fig. 2. An application model configures a generator orchestrator. The generator orchestrator uses the generator interface of each code generator to check for dependencies and computes an execution order. Finally, the generator orchestrator calls each code generator according to the computed execution order.

## 5 Realization in MontiArcAutomaton

The MontiArcAutomaton implementation of the conceptual model presented above comprises an implementation of generator interfaces, which is facilitated by a configuration language that generates interface implementations, an application configuration to declare compositions of code generators, and an orchestrator performing the composition.

### 5.1 Generator Interfaces in MontiArcAutomaton

Based on the concrete requirements for code generators in MontiArcAutomaton we refine the code generator interfaces defined in Def. 1. The C&C nature of MontiArcAutomaton, suggests separate interfaces for component generators and component behavior generators. As MontiArcAutomaton relies on factories for component and component behavior instantiation, factory generators are modeled as well. Component generators process the MontiArcAutomaton language and behavior generators process the respective embedded behavior modeling language (Def. 1, Item 1) possibly restricted by additional context conditions (Def. 1, Item 2). The classification in three generator kinds determines the output format of the generators (Def. 1, Item 3).

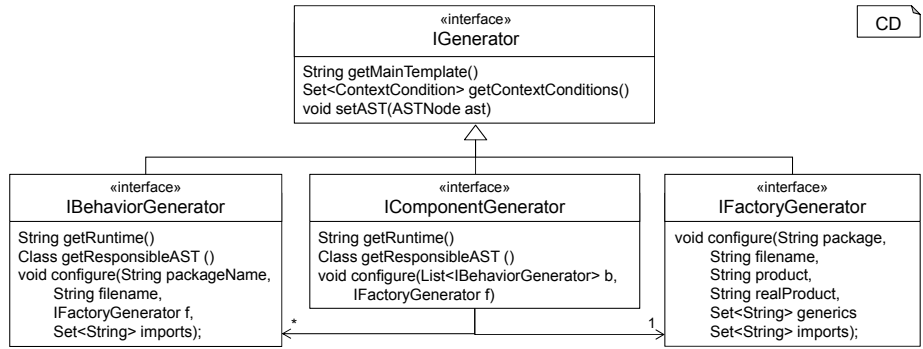

Fig. 3: Generator interface hierarchy of MontiArcAutomaton.

Generator execution information is provided by the generators in terms of the main template which the MontiCore code generation framework processes (Def. 1, Item 4). This template may call other templates and call Java code for complex calculations. All generators generate code conforming to a runtime environment they depend on (Def. 1, Item 5). The runtime environment determines, e.g., the scheduling of components. Generators in MontiArcAutomaton do not explicate further artifact dependencies as MontiArcAutomaton utilizes the delegator pattern [6] to integrate accordingly generated behavior implementations. Generation context information (Def. 1, Item 6) is provided to the generators at runtime and contains e.g., the AST of the processed model.

An overview of the concrete generator interfaces implemented for MontiArcAutomaton is displayed in Fig. 3. Every generator usable with MontiArcAutomaton implements an interface extending `IGenerator`. Thus, each generator can be parametrized with an AST node and provides at least its main template and its context conditions to the infrastructure. Generators for components and component behavior implement the interfaces `IComponentGenerator` and `IBehaviorGenerator` respectively. These interfaces explicate which AST types they can process.

Additionally, all generator interfaces define a method to `configure()` which is interface specific and defines the generation context information required. Generators for component behavior, e.g., expect to receive the package name of the containing component, the name of the artifact to be created, a factory generator, and the imported compilation units. The latter is required as embedding behavior into components produces integrated artifacts without distinction between the imports of the component and the imports of the behavior.

```
                                                    ┌──────────────────────────┐
                                                    │ GeneratorConfiguration   │
  ┌─────────────────────────────────────────────────┴──────────────────────────┴─┐
1 │ generator RobotArmPython {                                                     │
2 │     interface generators.IBehaviorGenerator;                                   │
3 │     template robotarm.Main;                                                    │
4 │     ast robotarm.ASTRobotArmProgram;                                           │
5 │     runtime runtimes.pythontimesync;                                           │
6 │ }                                                                              │
  └────────────────────────────────────────────────────────────────────────────┘
```

Listing 1: The generator configuration for the RobotArm generator describes that it implements the interface `IBehaviorGenerator` and provides static information.

## 5.2 Modeling Generator Interfaces

To facilitate the creation of code generator interfaces we have developed a modeling language for generator interfaces. Each code generator used with Monti­ArcAutomaton models how it is executed, which AST it processes, and which interface it implements in a single *generator configuration* model per generator. Listing 1 shows the model of the RA generator from the example in Sect. 3.1. This model describes that the generator implements the interface `IBehavior­Generator` and provides information accessible via this interface. The Monti­ArcAutomaton toolchain transforms these models into actual implementations implementing the interfaces.

The concrete implementation of the interface `IBehaviorGenerator` for the RobotArm generator from the example given in Lst. 1 provides implementations for all methods of `IGenerator` and `IBehaviorGenerator` and returns the static generator information from the model where applicable (e.g., `getRe­sponsibleAST()` returns an instance of the type specified behind `ast` in l. 4 of Lst. 1). The MontiArcAutomaton orchestrator can refer to these implementations via the implemented interfaces and compose generators as necessary.

## 5.3 Application Configuration and Generator Execution

Given a set of generators for component structure, behavior, and factories, an application has to specify which of these are to be used. This is modeled as the *application configuration* model. Listing 2 shows the application configuration for the toaster robot application. The model references a single component generator (l. 2), a single factory generator (l. 3), and two behavior generators - one for RA programs and one for I/O$^\omega$ automata (l. 4). An application configuration references at least a component structure generator and may reference additional behavior and factory generators.

Code generation in MontiArcAutomaton starts with the orchestrator processing the application configuration and loading the configuration of the referenced generators. As the order of generator execution is implicitly given by the C&C nature of MontiArcAutomaton, first the referenced behavior generators and the

```
                                                      ApplicationConfiguration
1  application ToasterRobotApplication {
2      componentgenerator ComponentsPython;
3      factorygenerator FactoryPython;
4      behaviorgenerators RobotArmPython, IOAutomatonPython;
5  }
```

Listing 2: Application configuration model for the toaster robot application using the RA generator for component behavior.

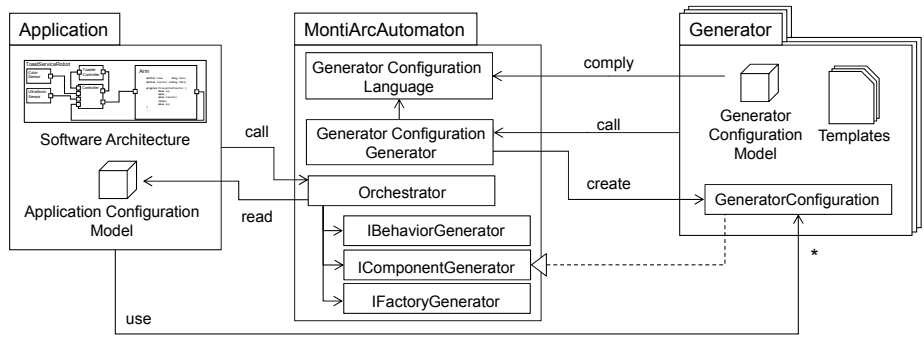

Fig. 4: Relations between applications using generators, the interfaces provided by MontiArcAutomaton, and the orchestrator performing the generator composition.

referenced factory generator are instantiated. Parametrized with these, the referenced component generator is instantiated. Afterwards, the orchestrator calls the main template of the component generator. The component generator traverses the AST of the architecture and thus also visits component behavior AST nodes. For each behavior AST node the responsible generator is configured with current AST generation context information and its main template is called with the AST node of the embedded behavior language.

Figure 4 shows the resulting relations: Applications consist of a software architecture, and an application configuration model. The application configuration model references the component, behavior, and factory generators required to build the software architecture of the project. To be processable by the orchestrator, referenced generators implement the appropriate generator interfaces.

## 6  Related Work

The presented approach for code generator composition is based on explicit generator interfaces, code generator orchestration, and application configuration. This approach is a first step towards a comprehensive approach for code generator composition and is closely related to modular code generator design.

The GenVoca model is an approach to build software systems generators based on composing object-oriented layers [1,2]. Different layers can use control blocks to exchange information. In contrast to this approach, we do not focus on a layered architecture of a code generator but an infrastructure for code generators composition.

The application building center is a multi-purpose modular framework for modeling software systems [18]. Genesys is an extension that allows to develop service-oriented code generators [9]. Each code generator represents a service that can be composed with other services. Information exchange is managed by using shared memory communication. Our presented approach is similar if we consider code generators to be services with interfaces. However, our approach introduces a broader generator interface to regard input language, output representation, input language constraints, execution information, artifact dependencies, and generation context information. This information is used to manage the execution and composition of the code generators.

Code generator composition using aspect-orientation at the artifact level has been described in [22]. The authors assume that a code generator produces operationally complete code fragments that are merged by a code fragment weaver. Additionally, in feature-oriented model-driven development (FO-MDD), multiple code generators are used to produce a software product line [20]. Composition of code is achieved after code generation by manually writing glue code. In contrast, we do not consider manual artifact composition but focus on an infrastructure to compose code generators. We nevertheless consider composing generated artifacts relevant for reusing code generators and will address this topic in future work.

## 7 Discussion and Future Work

We have presented a conceptual approach for composition of code generators based on the notion of generator interfaces. The ideas are implemented within the MontiArcAutomaton toolchain to enable post hoc embedding and use of new component behavior modeling languages. To broaden its applicability this approach requires future work on syntax, methods, and technical solutions.

Composition of arbitrary code generators without assumptions on their actual integration is harder to realize than for C&C ADLs. In general, generator composition demands a more expressive composition configuration than the application configuration presented above. For instance, the orchestration of the code generation process may require a code generator to be executed multiple times for every input model or to fill extension points provided by another generator under certain conditions. Moreover, execution of a code generator may not be triggered by a model type but by selecting a code generator for a particular set of input models. A generic model to configure an application has to express such process information and constraints. Thus, future research will look into modeling these aspects.

The generator composition illustrated above assumes that the orchestration of generators reflects the language embedding for component behavior. Other language integration mechanisms, such as language aggregation or language inheritance [11] will require a more complex orchestration. The generator for an inheriting language might, for example, require the generator for the inherited language to be executed first, such that the latter only generates additional artifacts for the model elements introduced by the inheriting language. Future work will therefore examine the notion of generator extension points as well.

Finally, modeling language composition mechanisms have lead to language reuse and language libraries. We hope to gain similar libraries and advantages from facilitating code generator composition.

## 8  Conclusion

We have motivated the need for generator composition in robotics and sketched a concept for code generator composition. This concept is based on explicit code generator interfaces and configuration models. The interfaces enable code generators to define information required for composition. A code generator orchestrator composes and executes the code generators. We have illustrated our implementation for the C&C modeling language family MontiArcAutomaton. Although the implementation relies on various assumptions implied by the language workbench MontiCore and the C&C nature of MontiArcAutomaton, we belief that these translate well into other contexts. There are however open issues in arbitrary generator composition and we have identified possible extensions of generator interfaces and generator orchestrators to be applied in more complex scenarios.